\pdfoutput=1

\documentclass[11pt,a4paper]{article}
\usepackage[hyperref]{acl2017}

\aclfinalcopy 
\def\aclpaperid{RepLNLP36} 
\usepackage{cmap} 
\usepackage[english]{babel}
\usepackage[utf8]{inputenc}
\usepackage{amsmath,amssymb}
\usepackage{txfonts}
\usepackage[T1]{fontenc}
\usepackage{graphicx}
\usepackage{xcolor}
\usepackage{xparse}
\usepackage{varioref}
\usepackage{subcaption}
\usepackage[normalem]{ulem}

\usepackage[protrusion,expansion]{microtype} 
\usepackage{booktabs}	
\usepackage{xspace}	

\newcommand{\orcid}[1]{\href{https://orcid.org/#1}{\texttt{#1}}}
\newcommand{\email}[1]{\href{mailto:#1}{\texttt{#1}}}
\newcommand*\rot{\rotatebox{65}}
\newcommand*\rotw{\rotatebox{35}}

\newcommand{\Ocls}[1]{\ensuremath{\mathcal{O}(#1)}}

\newcommand{\ES}{Elasticsearch\xspace}

\newcommand{\Luc}{Lucene\xspace}
\newcommand{\op}[1]{\ensuremath{\operatorname{#1}}}

\newcommand\cipherphantom{\hphantom{0}}
\def\NA{all\xspace}
\let\stress\emph
\long\def\motto#1\par#2\\#3 {}

\newcommand\Authors{Jan Rygl; Jan Pomikálek; Radim Řehůřek;
                    Michal Růžička; Vít Novotný; Petr Sojka}
\newcommand\Title{Semantic Vector Encoding and Similarity Search Using~Fulltext~Search~Engines}
\newcommand{\Subject}{RepL4NLP@ACL 2017 paper}
\newcommand{\Keywords}{%
	vector space modelling;
	semantic vectors encodings;
	inverted-index-based systems performance;
	document feature representations;
	search efficiency;
	Latent Semantic Analysis;
	Elasticsearch;
	curse of dimensionality;
	information system performance optimization}
\title{\Title}
\author{%
  Jan Rygl \and Jan Pomikálek \and\\
   \textbf{Radim {\v R}eh{\r u}{\v r}ek} \\
  RaRe Technologies \\
  {\tt \email{jimmy@rare-technologies.com}} \\
  {\tt \email{honza@rare-technologies.com}} \\
  {\tt \email{radim@rare-technologies.com}} \\
  \qquad{}\And{}\qquad
  Michal R{\r u}{\v z}i{\v c}ka \and V{\'i}t Novotn{\'y}\and\\ 
  \textbf{Petr Sojka} \\
  Faculty of Informatics, Masaryk University \\
  Botanick\'a 68a, 602\,00 Brno, Czechia \\
  {\tt \email{mruzicka@mail.muni.cz}}, \\
  {\tt ~~\email{witiko@mail.muni.cz}}, \\
  {\tt ~\email{sojka@fi.muni.cz}}\phantom,\\
  \small ORCID:  \orcid{0000-0001-5547-8720},\\
  \small \orcid{0000-0002-3303-4130},
                \orcid{0000-0002-5768-4007}
}

\let\vref\ref 

\begin{document}
\hyphenation{Elastic-search} 
\hypersetup{%
  pdftitle={\Title},
  pdfauthor={\Authors},
  pdfkeywords={\Keywords},
  pdfsubject={\Subject}}

\maketitle

\vspace*{-3\baselineskip}
\begin{abstract}
Vector representations and vector space modeling (VSM) play a central role in modern machine learning. We propose a novel approach to `vector similarity searching' over dense semantic representations of words and documents that can be deployed on top of traditional inverted-index-based fulltext engines, taking advantage of their robustness, stability, scalability and ubiquity.
We show that this approach allows the indexing and querying of dense vectors in text domains. This opens up exciting avenues for major efficiency gains, along with simpler deployment, scaling and monitoring.
The end result is a fast and scalable vector database with a tunable trade-off between vector search performance and quality, backed by a standard fulltext engine such as \ES.
We empirically demonstrate its querying performance and quality by applying this solution
to the task of semantic searching over a dense vector representation of the entire English Wikipedia.
\end{abstract}

\section{Introduction}
\label{sec:intro}

The vector space model~\cite{ir:Salton1975} of representing documents in high-dimensional vector spaces has been validated by decades of research and development. Extensive deployment of inverted-index-based information retrieval (IR) systems has led to the availability of robust open source IR systems such as Sphinx, Lucene or its popular, horizontally scalable extensions of Elasticsearch and Solr.

Representations of document semantics based solely on 
first order document-term statistics, such as TF-IDF
or \href{https://en.wikipedia.org/wiki/Okapi_BM25}{Okapi BM25}, are 
limited in their expressiveness and search recall. Today, approaches
based on distributional semantics and deep learning allow the construction
of semantic vector space models representing words, sentences,
paragraphs or even whole documents as vectors in high-dimensional
spaces~\cite{nlp:LSA1990,blei03lda,ml:mikolov2013distributed}.

The ubiquity of semantic vector space modeling raises the challenge of 
efficient searching in these dense, high-dimensional vector spaces.
We would naturally want to take advantage of the design and optimizations behind modern fulltext engines like Elasticsearch so as to meet the scalability and robustness demands of modern IR applications. This is the research challenge addressed in this paper.

The rest of the paper describes novel ways of encoding dense vectors into text documents, allowing the use of traditional inverted index engines, and explores the trade-offs between IR accuracy and speed. Being motivated by pragmatic needs, we describe the results of experiments carried out on real datasets measured on concrete, practical software implementations.

\section{Semantic Vector Encoding for~Inverted-Index Search Engines}

\subsection{Related Work}
The standard representation of documents in the Vector Space Model (VSM)~\cite{ml:SaltonBuckley1988}
uses term feature vectors of very high dimensionality.
To map the feature space onto a smaller and denser latent semantic subspace, we may use a body of techniques, including Latent Semantic Analysis (LSA)~\cite{nlp:LSA1990}, Latent Dirichlet Allocation (LDA)~\cite{blei03lda} or the many variants of  Locality-sensitive hashing (LSH)~\cite{ir:conf/vldb/GionisIM99}. 

Throughout the long history of VSM developments, many other methods for improving search efficiency have been explored. \citeauthor{ir:Weber1998} ran one of the first rigorous studies that dealt with the ineffectiveness of the VSM and the so-called curse of dimensionality. 
They evaluated several data partitioning and vector approximation schemes, achieving significant nearest-neighbour search speedup in ~\cite{ir:Weber2000}. 
The scalability of similarity searching through a  new index data structures design is described in~\cite{ir:Zezula2006}.  
Dimensionality reduction techniques proposed in~\cite{ir:Digout2004} allow a robust speedup 
while showing that not all features are equally discriminative and have a different impact on efficiency, due to their density distribution. 
\citeauthor{ir:boytsov2017} shows that the $k$-NN search can be a replacement for term-based retrieval when the term-document similarity matrix is dense.

Recently, deep learning approaches and tools like doc2vec~\cite{ml:LeMikolov2014} construct \stress{semantic} representations of documents~$D$ as $d=|D|$ (dense) vectors in an $n$-dimensional vector space.

To take advantage of ready-to-use and optimized systems for term indexing and searching, we have developed a method for representing points in a semantic vector space encoded as plain text strings. In our experiments, we will be using \ES~\cite{Gormley:2015:EDG:2904394} in its `vanilla' setup. We do not utilize any advanced features of \ES, such as custom scoring, tokenization or a $n$-gram analyzers. 
Thus, our method does not depend on any functionality that is specific to \ES, and it is possible (and sometimes even desirable) to substitute \ES with other fulltext engine implementations.

\subsection{Our Vector to String Encoding Method}
\label{sec:encoding}
Let our query be a document, represented by its vector $\vec q$, for which we aim to find the top~$k$ most similar documents in~$D$.
We want to search efficiently, indexing and deleting documents from the index in near-real-time, and in a manner that could scale by eventual parallelization, re-using the significant research and engineering effort that went into designing and implementing systems like \ES.

Conceptually, our method consists of encoding vector features into string tokens (feature tokens), creating a text document from each dense vector. These encoded documents are consequently indexed in traditional inverted-index-based search engines. At query time, we encode the query vector and retrieve the subset of similar vectors~$E$, $|E| \ll |D|$, using the engine's fulltext search functionality.
Finally, the small set of candidate vectors~$E$ is re-ranked by calculating the exact similarity metric (such as cosine similarity) to the query vector. This makes the search effectively a \stress{two-phase process}, with our encoded fulltext search as the first phase and candidate re-ranking as the second.

\subsubsection{Encoding Vectors}
\label{sec:stringification}

The core of our method of encoding vectors into strings lies in encoding the vector feature values at a \stress{selected precision} as character strings~-- feature tokens.%
\footnote{We avoid the use of any special characters such as plus ($+$) or 
          minus ($-$) signs, white spaces etc. inside the tokens. This is used as 
          a safeguard against any unintended tokenization within the fulltext search system, 
          so as to avoid having to define custom tokenizers.}
This is best demonstrated on a small example:
Let us take a semantic vector of three dimensions,
$\vec w  = [0.12, -0.13, 0.065]$.
Each feature token starts with its feature identification (e.g.\ a feature number such as \texttt{0}, \texttt{1} etc.) followed by a precision 
encoding schema identifier (such as \texttt{P2}, \texttt{I10} etc.) and the encoded feature value (such as \texttt{i0d12}, \texttt{ineg0d2} etc.) depending on the particular encoding method. We propose and evaluate three encoding methods:

\begin{description}
    \item[rounding] 
    Feature values are rounded to a fixed number of decimal 
    positions and stored as a string that encodes both the feature identification
    and its value. Rounding to two decimal places produces representation 
    of $\vec w$ as [\verb|'0P2i0d12'|, \verb|'1P2ineg0d13'|, \verb|'2P2i0d07'|].
    
    \item[interval] quantizes $\vec w$ into intervals of fixed length. For example, with an interval width of $0.1$, feature values fall into intervals starting at 0.1, $-0.2$ and 0.0, which we encode as \texttt{d1}, \texttt{d2} and \texttt{d0}, respectively. Combined with the interval length denotation of \texttt{I10}, the full vector is encoded into the tokens
    $\vec w= [$\verb|'0I10i0d1'|, \verb|'1I10ineg0d2'|, \verb|'2I10i0d0'|].

    \item[combined] Rounding and interval encoding used together.
    Rounding $\vec w$ to three decimal places and using intervals of length $0.2$, we get
    the representation of $\vec w$ as feature tokens
    [\verb|'0P3i0d120'|, \verb|'1P3ineg0d130'|, \verb|'2P3i0d065'|,
    \verb|'0I5i0d0'|, \verb|'1I5ineg0d2'|, \verb|'2I5i0d0'|].
\end{description}

The intuition behind all these encoding schemes is a trade-off between \textit{increasing feature sparsity} and \textit{retaining search quality}: we show that some types of sparsification actually lead to little loss of quality, allowing an efficient use of inverted-index IR engines.

\subsubsection{High-Pass Filtering Techniques}
\label{sec:hp-filtering}

The rationale behind the next two techniques is to filter out semantic vector features of low importance, further increasing feature sparsity. This improves performance at the expense of quality.

\paragraph{Trim:} In the trimming phase, a fixed threshold~-- such as 0.1~-- is used.
Feature tokens in the query with an absolute value of the feature below the threshold are simply discarded from the query.
In the case of our example vector $\vec w  = [0.12, -0.13, 0.065]$, the tokens representing the third feature value 0.065 are removed since $|0.065| < 0.1$.

\paragraph{Best:} Features of each vector are ordered according to their absolute value and only a fixed number of the highest-valued features are added to the index, discarding the rest. As an example, with $best=1$, only the second feature of $-0.13$ (the highest absolute value) would be considered from~$\vec w$.

Note that in both cases, this type of filtering is only meaningful when the feature ranges are comparable. In our experiments all vectors are normalized to unit length, ranging absolute values of features from zero (no feature importance) to one (maximal feature importance). 

\subsection{Space and Time Requirements}
\label{sec:space-time-reqs}
In this section we summarize and compare the theoretical space and time
requirements of our proposed vector-to-string encoding and filtering methods with a baseline of a naive, linear brute force search.
The inverted-index-based analysis is based on the documentation of the \Luc search engine implementation\footnote{\url{https://lucene.apache.org/core/6_5_1/index.html}}.

While the running time of the naive search is stable and predictable,
the efficiency of the other optimization methods depends 
on the data set, such as its vocabulary size and the expected postings list sparsity, after the feature token encoding and filtering.
On the other hand, performance can be influenced by how the method is configurated  -- for example, the expected number of distinct feature values depends on the precision of the rounding of the feature values that was used. 

The efficiency trade-offs are summarized in Table~\vref{tab:methods-estimates}.
Each document is represented as a vector $\vec{d}$ of $n$~features computed by LSA over TF-IDF.

\paragraph{Baseline~-- naive brute force search:}
The naive baseline method avoids using a fulltext search altogether, and instead stores all~$d$ `indexed' vectors in their original dense vector representation in RAM, as an $n \times d$ matrix.
At query time, it computes a similarity between the query vector~$\vec q$ and each of the indexed document vectors~$\vec d$, in a single linear scan through the matrix, and returns the top $k$ vectors with the best score.

\paragraph{Efficiency of a naive brute force search:}
The index is a matrix of floats of size $n d$ resulting in $\Ocls{n d}$ space 
complexity.  We have chosen the cosine similarity as our similarity metric.
To calculate cossim between the query and all $d$~documents, 
vector $\vec q$ of length~$n$ has to be multiplied with a length-normalized matrix of dimensionality $n\times d$, e.g.\ $\Ocls{n d}$. Using the resulting vector of $d$~scores, we then pick the $k$~nearest documents as the final query result, in~$\Ocls{d}$.

\paragraph{Efficiency of encoding:} 
We investigate the efficiency of our vector-to-string encoding when
a general inverted-index-based search engine is used to index and search through
them.

At worst, we store one token per dimension for each vector.
We need $\Ocls{n d}$ space to store all indexed documents, as is the case with the naive search.
In practice, there are several different constants as naive search saves 
one float per feature (four bytes), while our feature tokens are compressed string-encoded feature values and indices of the sparse feature positions in the inverted index.

Each dimension of the query vector $\vec{q}$ contains a string-encoded feature $q_j$.
For each $q_j$ we fetch a list of documents $c_i$ together with
term frequency $t_n$ in that document: $\op{tf}(t_n, c_i)$.

For each of these $(c_i, \op{tf}(t_n, c_i))$ pairs we add 
the corresponding score value to the score of document $c_i$ in 
a list of result-candidate documents. The score computation 
contains a vector dot-product operation in the dimension of the 
size~$v$ of feature vocabulary $V$ that can be computed in $\Ocls{v}$. 
Document~$c_i$ is added to the set~$C$ of all result-candidate 
documents, which we sort in $\Ocls{c \log{c}}$ time
and return the top $k$~results. The whole search is performed
in $\Ocls{n \cdot p \cdot v + c\log{c}}$ steps, where $p$ is the 
expected postings list size and $c = |C|$. 

\paragraph{Efficiency of high-pass filtering:}
We approximate the full feature vector by storing only the most significant features, e.g.\ only the top $m$~dimensions. When compared with a naive search, we save on space: only $\Ocls{m d}$, $m\ll n$ values are needed.

For each feature value $q_j$ we have to find 
all documents with the same feature value.
We are able to find the feature set in $\Ocls{\log{j}}$ steps, where 
$j$~is the number of distinct indexed values of the feature.
Consequently, we retrieve the matched documents in 
$\Ocls{l}$ time, where $l$~is the number of documents in the index with the 
appropriate feature value.

Each of the found documents is added to set~$C$ and its score is 
incremented for this hit. If represented with an appropriate data structure,
such as a hash table, we are able to add and increment scores of the items in $\Ocls{1}$ time.
Having all $c = |C|$~candidate documents over all the separate feature-searches, we pick and return the top $k$~items in $\Ocls{c}$ time.

Combined, $\Ocls{n \cdot (\log{j} + l) + c}$ steps are needed
for the search, where $j$ is the expected number of distinct
indexed values per feature and $l$ is the number of documents in the
index per feature value.

\begin{table*}[htb]
  \tabcolsep4dd\centering
  \caption{Comparison of encoding methods in terms of space and time.
     $n$~is the number of semantic vector features,
     $d$~is the number of semantic vectors,
     $m$~is the number of semantic vector features after high-pass filtering,
     $p$~is the expected postings list size (inverted index sparsity),
     $v$~is the expected vocabulary size per a feature,
     $c$~is the number of result-candidate semantic vectors,
     $j$~the expected number of distinct indexed values of the feature, 
         and
     $l$~the expected number of documents in the index per a feature value.
     For details see Section~\vref{sec:space-time-reqs}.}
  \label{tab:methods-estimates}
  \begin{tabular}{lccc}
    \toprule
             & naive search & token encoding & high-pass filtering\\
    \midrule
       space & $\Ocls{n d}$ & $\Ocls{n d}$   & $\Ocls{m d}$   \\
        time & $\Ocls{n d}$ & $\Ocls{n \cdot p \cdot v + c\log{c}}$  & $\Ocls{n \cdot (\log{j} + l) + c}$ \\
    \bottomrule
  \end{tabular}
\end{table*}

\section{Experimental Setup}
To evaluate our method, we used ScaleText~\cite{ir:raslanscaletext2016} based on \ES~\cite{Gormley:2015:EDG:2904394} as our fulltext IR system.
The evaluation dataset was the whole of the English Wikipedia consisting of 4,181,352 articles.

\subsection{Quality Evaluation}
\label{sec:quality-eval}
The aim of the quality evaluation was to investigate how well the approximate `encoded vector' search performs in comparison with the exact naive brute-force search, using cosine similarity as the similarity metric. Cosine similarity is definitely not the only possible metric~-- we selected cosine similarity as we needed a fully automatic evaluation, without any need of human judgement, and because cosine similarity suits our upstream application logic perfectly.

We converted all Wikipedia documents into vectors using LSA with 400 features.
We then randomly selected 1,000 documents from our Wikipedia dataset to act as our query vectors.
By doing a naive brute force scan over all the vectors (the whole dataset), we identified the 10~most similar ones for each query vector. This became our `gold standard'.

We encoded the dataset vectors into various string representations, as described in Section~\vref{sec:stringification} and stored them in \ES.

For evaluating the search, we pruned the values in the query vectors (see Section~\vref{sec:hp-filtering}) and encoded them into a string representation. Using these strings, we performed 1,000 \ES searches. For each query, we measured the overlap between the retrieved documents and the gold standard using Precision@k or the Normalized Discounted Cumulative Gain (nDCG$_k$). The mean cumulative loss between the ideal and the actual cosine similarities of the top~$k$ results (avg.\ diff.) is also reported.

Note that since we re-rank $|E|$ results obtained from \ES (see Section~\vref{sec:encoding}), the positions on which the gold standard vectors were originally returned by the fulltext engine are irrelevant.

Apart from the vector dimensionality $n$ (the number of LSA \textbf{features}), we monitored the \textbf{trim} threshold and the number of \textbf{best} features as described in Section~\vref{sec:hp-filtering}.
We also experimented with the number of vectors $E$ retrieved for each \ES{} query as the \textbf{page} parameter.

To provide a comparison with an established search approach, and to serve as a baseline, we also evaluated indexing and searching using the native fulltext indexing and searching capabilities of \ES.
In this case, the plain fulltext of every article was sent directly to the 
fulltext search engine as a string, without any vector conversions or preprocessing.
For querying, we use the \stress{More Like This (MLT) Query} API of \ES.%
\footnote{\url{https://www.elastic.co/guide/en/elasticsearch/reference/5.2/query-dsl-mlt-query.html}}
Any data processing during indexing and querying was done by \ES in its default settings
with a single exception: we evaluated multiple values of the 
\verb|max_query_terms| parameter of the MLT API. We tested the default value 
(25) plus values corresponding to the values of the \textbf{best}
parameter used for the evaluation of our method.

We report mainly on the \textit{avg.\ diff.\@}, i.e.\ the mean difference between the ideal and the actual cosine similarities of the first ten retrieved documents to a query, averaged over all 1,000 queries. We also report \textit{Precision@10}, i.e.\ the ratio of the gold standard documents in the first ten results averaged over all 1,000 queries, and on \textit{Normalized Discounted Cumulative Gain (nDCG$_{10}$)}~\cite[Section~8.4]{ir:Manning:2008:IIR}, where the relevance value of a retrieved document is taken to be its cosine similarity to the query.

\subsection{Speed Evaluation}
In this section, we evaluate the performance of feature token strings searches in \ES{}, using various \ES{} configurations as well as various vector filtering parameters.

Our \ES{} cluster consisted of 6~nodes, with 32\,GiB RAM and 8~CPUs each, for a total of 192\,GiB and 48~cores.

We experimented with several \ES{} parameters: the number of \ES{} index shards (6, 
12, 24, 48, 96; always using one replica), the number of LSA features (100, 200, 
400), a trim threshold of LSA vector values (none, 0.05, 0.10, 0.20), the number of \ES{} 
results used (\ES{} page size; 20, 80, 320, 640), parallel querying (1 [serial], 4, 
16) and the cluster querying strategy (querying single-node or round-robin querying 
of 5~different \ES{} nodes).
Each evaluation batch consisted of 128~queries (randomly selected for each batch) that were used to ask \ES one-by-one or in parallel in multiple queues depending on the evaluation setup.

We report \textit{ES avg./std.\,[s]}~-- the average number of seconds per request \ES took (the `took' time from the \ES response, i.e.\ the search time inside \ES) and its standard deviation, \textit{Request avg./std.\,[s]}~-- the average number of seconds and its standard deviation per request including communication overhead with \ES to get the results (i.e.\ the time of the client to get the answer), 
\textit{Total time [s]}~-- total number of seconds for processing all requests in the batch. This can differ from the sum of average request times when executing queries in parallel.
The number of features in query vectors that passed through threshold trimming is reported as \textit{Vec.\ size avg./std.}

\section{Results}

\subsection{Quality Evaluation}
\label{sec:eval:quality}

Results of the quality evaluation are summarized in Table~\vref{tab:results:quality}.
The results of our method in different settings are put side by side with the 
results of the brute-force naive search and with \ES's native More Like 
This (MLT) search. For the MLT results, the 
\verb|max_query_terms| \ES{} parameter is reported in the \textbf{best} column
since both the semantics and the impact on the search speed are similar.

\begin{table*}[p]
    \renewcommand{\arraystretch}{1.22}
    \tabcolsep2.8dd
\def\prec.{prec.} \def\prec.{P@10} \caption{Results of quality evaluation. 400 
LSA features were used.
         See Section~\vref{sec:quality-eval} for more details.
         Only the subset of results with the top nDCG$_{10}$ scores are shown.
         Results with Precision@10 $\geq 0.9$ are shown in bold,
         with avg. diff. $\leq 0.002$ in italics.}
\vspace*{-.25\baselineskip}
\small
\bgroup\let~\cipherphantom
\def\a{\bf\em}
\def\b{\bf}
\def\i{\it}
\begin{tabular}[t]{@{}c c c c c c c c}
  \toprule
  \rot{Trim} & \rot{Best} & \rot{Page} & \rot{Min.\ \prec.} & \rot{Avg.\ \prec.} & \rot{Max.\ \prec.} & \rot{nDCG$_{10}$} & \rot{Avg.\ diff.} \\
  \midrule
  \multicolumn{2}{l }{\a naive search}%
            &\a ~10 &\a      1.0 &\a   1.0000 &\a      1.0 &\a1.0000 &\a  0.0000 \\
  \midrule
  MLT  & ~17 &    10 &        0.0 &     0.1967 &        1.0 &  0.9206 &    0.1799 \\
  MLT  & ~25 &    10 &        0.0 &     0.2029 &        1.0 &  0.9221 &    0.1698 \\
  MLT  & ~40 &    10 &        0.0 &     0.2077 &        1.0 &  0.9224 &    0.1619 \\
  MLT  & ~90 &    10 &        0.0 &     0.2120 &        1.0 &  0.9210 &    0.1580 \\
  MLT  & 400 &    10 &        0.0 &     0.2114 &        1.0 &  0.9211 &    0.1568 \\
  \midrule
  0.00 &  160 &   17 &        0.0 &     0.4275 &        1.0 &  0.9840 &    0.0220 \\
  0.00 &  320 &   17 &        0.0 &     0.5281 &        1.0 &  0.9949 &    0.0149 \\
  0.00 &  640 &   17 &        0.0 &     0.6340 &        1.0 &  0.9989 &    0.0101 \\
  0.00 &   40 &   40 &        0.0 &     0.5090 &        1.0 &  0.9870 &    0.0142 \\
  0.00 &   80 &   40 &        0.0 &     0.6215 &        1.0 &  0.9940 &    0.0085 \\
  0.00 &  160 &   40 &        0.0 &     0.7254 &        1.0 &  0.9980 &    0.0051 \\
  0.00 &  320 &   40 &        0.1 &     0.8181 &        1.0 &  0.9998 &    0.0030 \\
\i0.00 &\i640 &\i 40 &\i      0.0 &\i   0.8883 &\i      1.0 &\i0.9989 &\i  0.0016 \\
  0.00 &   10 &   90 &        0.0 &     0.3927 &        1.0 &  0.9940 &    0.0323 \\
  0.00 &   20 &   90 &        0.0 &     0.5167 &        1.0 &  0.9950 &    0.0147 \\
  0.00 &   40 &   90 &        0.0 &     0.6314 &        1.0 &  0.9970 &    0.0084 \\
  0.00 &   80 &   90 &        0.1 &     0.7306 &        1.0 &  0.9994 &    0.0051 \\
  0.00 &  160 &   90 &        0.1 &     0.8185 &        1.0 &  0.9994 &    0.0029 \\
\i0.00 &\i320 &\i 90 &\i      0.0 &\i   0.8818 &\i      1.0 &\i0.9988 &\i  0.0017 \\
\a0.00 &\a640 &\a 90 &\a      0.0 &\a   0.9282 &\a      1.0 &\a0.9989 &\a  0.0010 \\
  0.00 &   20 &  \NA &        0.1 &     0.5730 &        1.0 &  1.0000 &    0.0124 \\
  0.00 &   80 &  \NA &        0.3 &     0.7870 &        1.0 &  1.0000 &    0.0044 \\
\a0.00 &\a320 &\a\NA &\a      0.4 &\a   0.9050 &\a      1.0 &\a1.0000 &\a  0.0016 \\
\a0.00 &\a640 &\a\NA &\a      0.4 &\a   0.9460 &\a      1.0 &\a1.0000 &\a  0.0010 \\
  0.05 &  160 &   17 &        0.0 &     0.4276 &        1.0 &  0.9837 &    0.0220 \\
  0.05 &  320 &   17 &        0.0 &     0.5280 &        1.0 &  0.9948 &    0.0149 \\
  0.05 &  640 &   17 &        0.0 &     0.6341 &        1.0 &  0.9989 &    0.0101 \\
\bottomrule
\end{tabular}\hfill\begin{tabular}[t]{c c c c c c c c@{}}
  \toprule
  \rot{Trim} & \rot{Best} & \rot{Page} & \rot{Min.\ \prec.} & \rot{Avg.\ \prec.} & \rot{Max.\ \prec.} & \rot{nDCG$_{10}$} & \rot{Avg.\ diff.} \\
  \midrule
  0.05 &   40 &   40 &        0.0 &     0.5088 &        1.0 &  0.9870 &    0.0142 \\
  0.05 &   80 &   40 &        0.0 &     0.6216 &        1.0 &  0.9940 &    0.0085 \\
  0.05 &  160 &   40 &        0.0 &     0.7255 &        1.0 &  0.9980 &    0.0051 \\
  0.05 &  320 &   40 &        0.1 &     0.8177 &        1.0 &  0.9992 &    0.0030 \\
\i0.05 &\i640 &\i 40 &\i      0.0 &\i   0.8880 &\i      1.0 &\i0.9988 &\i  0.0016 \\
  0.05 &   10 &   90 &        0.0 &     0.3927 &        1.0 &  0.9940 &    0.0323 \\
  0.05 &   20 &   90 &        0.0 &     0.5168 &        1.0 &  0.9950 &    0.0147 \\
  0.05 &   40 &   90 &        0.0 &     0.6313 &        1.0 &  0.9970 &    0.0084 \\
  0.05 &   80 &   90 &        0.0 &     0.7305 &        1.0 &  0.9990 &    0.0051 \\
  0.05 &  160 &   90 &        0.1 &     0.8187 &        1.0 &  0.9997 &    0.0029 \\
\i0.05 &\i320 &\i 90 &\i      0.0 &\i   0.8815 &\i      1.0 &\i0.9988 &\i  0.0017 \\
\a0.05 &\a640 &\a 90 &\a      0.0 &\a   0.9281 &\a      1.0 &\a0.9988 &\a  0.0010 \\
  0.05 &   10 &  \NA &        0.0 &     0.3923 &        1.0 &  0.9960 &    0.0321 \\
  0.05 &   20 &  \NA &        0.0 &     0.5154 &        1.0 &  0.9976 &    0.0149 \\
  0.05 &   40 &  \NA &        0.0 &     0.6320 &        1.0 &  0.9985 &    0.0085 \\
  0.05 &   80 &  \NA &        0.0 &     0.7321 &        1.0 &  0.9990 &    0.0051 \\
  0.05 &  160 &  \NA &        0.0 &     0.8179 &        1.0 &  0.9990 &    0.0030 \\
\i0.05 &\i320 &\i\NA &\i      0.1 &\i   0.8810 &\i      1.0 &\i0.9992 &\i  0.0018 \\
\a0.05 &\a640 &\a\NA &\a      0.1 &\a   0.9302 &\a      1.0 &\a0.9991 &\a  0.0009 \\
  0.10 &  320 &   17 &        0.0 &     0.4981 &        1.0 &  0.9888 &    0.0171 \\
  0.10 &  640 &   17 &        0.0 &     0.6008 &        1.0 &  0.9969 &    0.0117 \\
  0.10 &  160 &   40 &        0.0 &     0.4409 &        1.0 &  0.9771 &    0.0216 \\
  0.10 &  320 &   40 &        0.0 &     0.5432 &        1.0 &  0.9891 &    0.0148 \\
  0.10 &  640 &   40 &        0.0 &     0.6435 &        1.0 &  0.9968 &    0.0099 \\
  0.10 &  320 &   90 &        0.0 &     0.5434 &        1.0 &  0.9893 &    0.0148 \\
  0.10 &  640 &   90 &        0.0 &     0.6436 &        1.0 &  0.9968 &    0.0099 \\
  0.10 &  160 &  \NA &        0.0 &     0.4410 &        1.0 &  0.9770 &    0.0216 \\
  0.10 &  320 &  \NA &        0.0 &     0.5431 &        1.0 &  0.9889 &    0.0148 \\
  0.10 &  640 &  \NA &        0.0 &     0.6438 &        1.0 &  0.9972 &    0.0099 \\
    \bottomrule
\end{tabular}
\egroup
\label{tab:results:quality}
\vspace{2em}
\end{table*}

\begin{figure*}[p]
    \begin{subfigure}[t]{.49\textwidth}
        \includegraphics[width=1.1\textwidth]{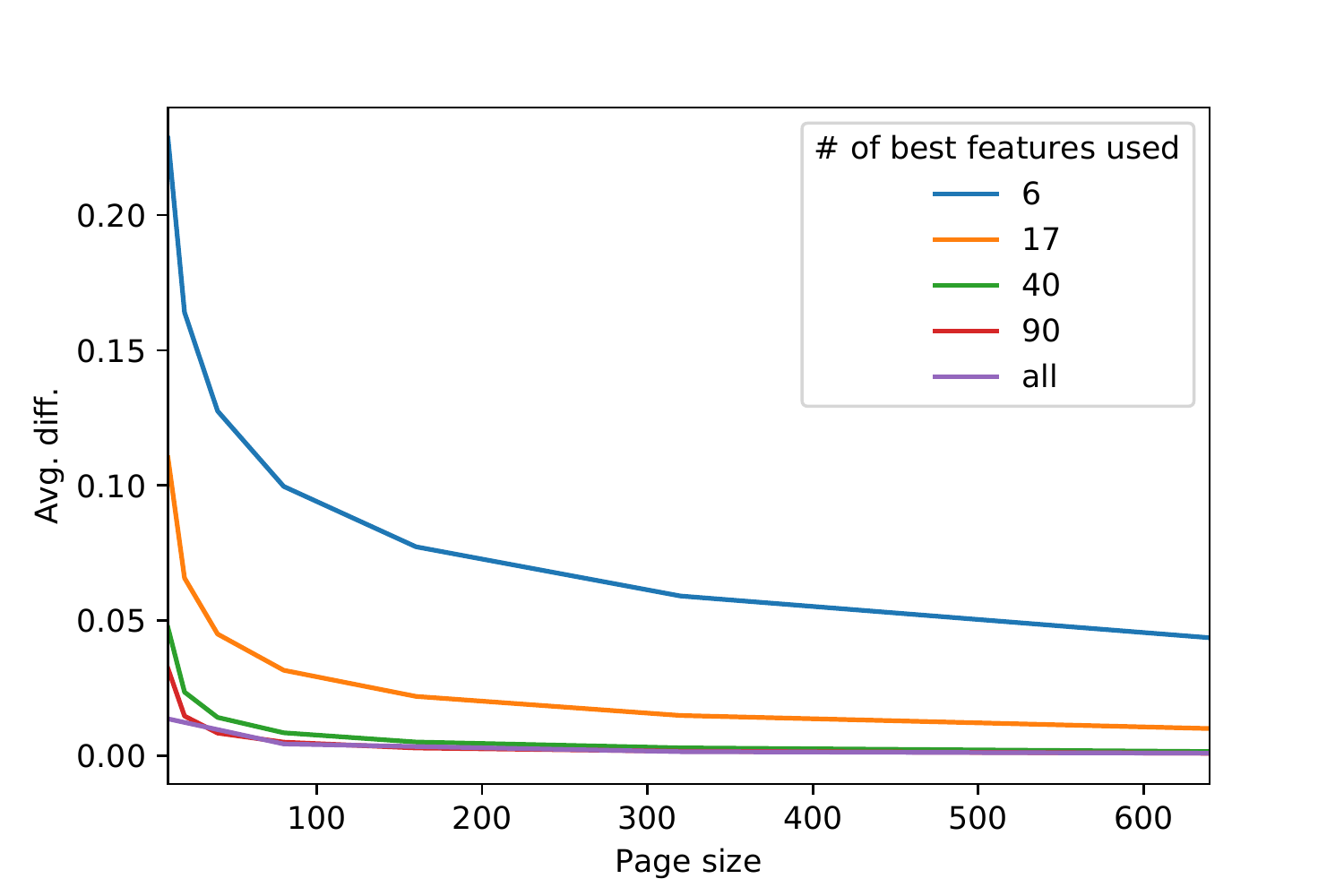}
        \caption{Avg.\ diff.}
        \label{fig:results:quality:avg-diff}
    \end{subfigure}
    \hfill
    \begin{subfigure}[t]{.49\textwidth}
        \includegraphics[width=1.1\textwidth]{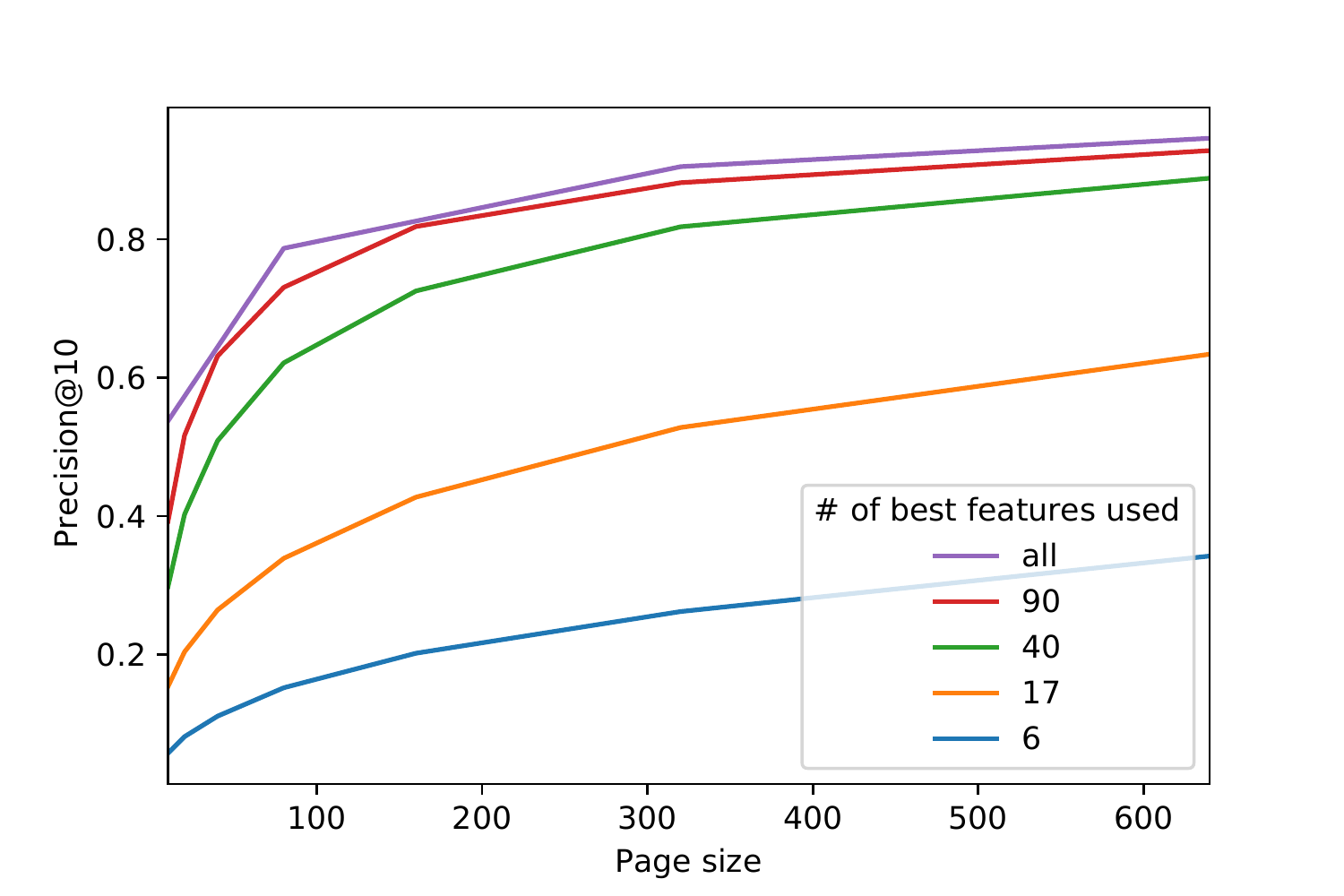}
        \caption{Precision@10}
        \label{fig:results:quality:precision}
    \end{subfigure}
    \caption{The impact of
             the number of best features selected (with no trimming) and
             the page size (the number of search results retrieved from \ES)
             on Avg.\ diff.\ and Precision@10.}
    \label{fig:results:quality}
\end{figure*}

\begin{figure*}[tb]
    \begin{subfigure}[t]{0.499\textwidth}
        \captionsetup{justification=raggedright}
        \centerline{\includegraphics[width=.95\textwidth]{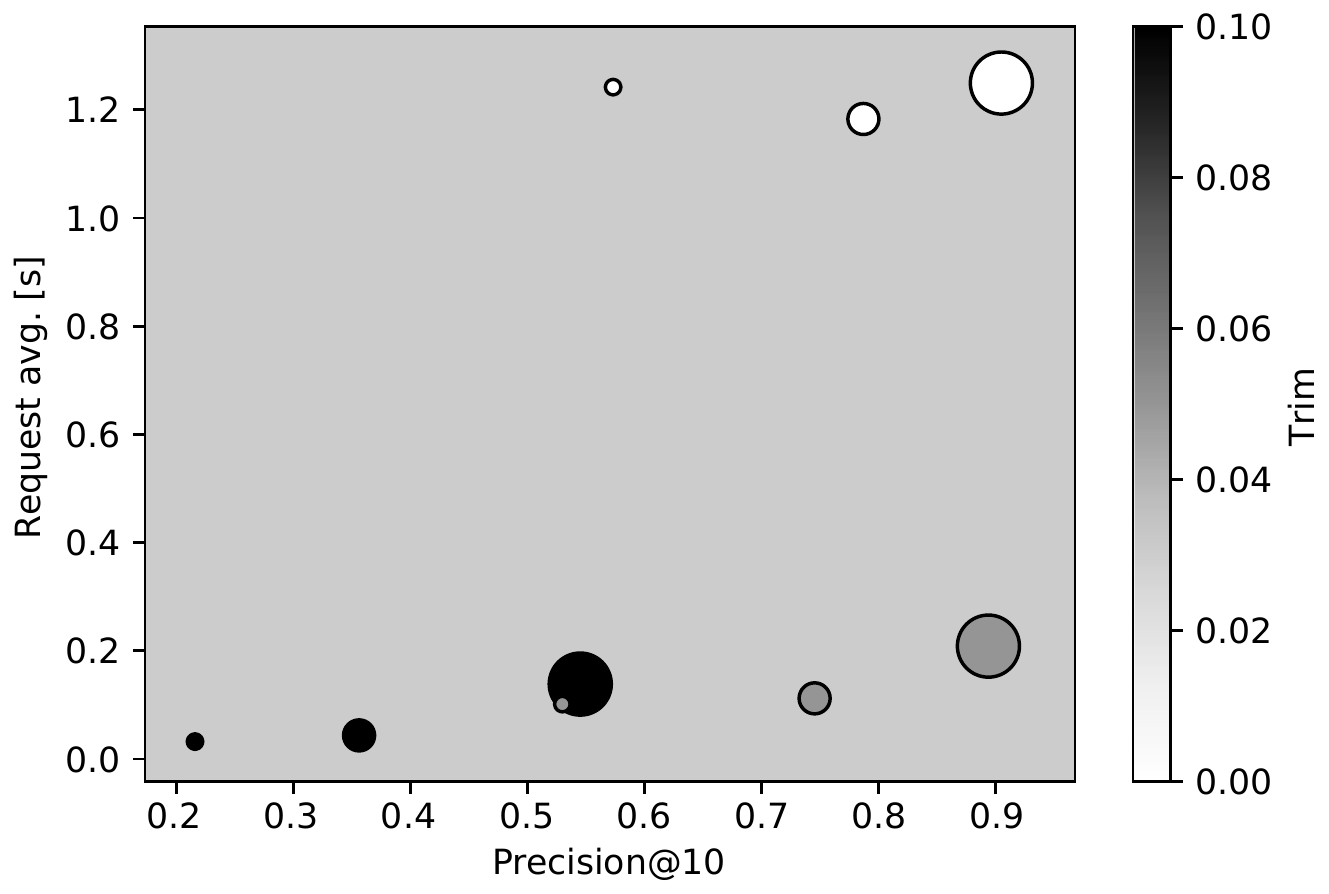}}
    \caption{Consecutive querying of 128~queries}
        \label{fig:results:req-time-avg:single}
    \end{subfigure}
    \hfill
    \begin{subfigure}[t]{0.499\textwidth}
        \captionsetup{justification=raggedright}
        \centerline{\includegraphics[width=.95\textwidth]{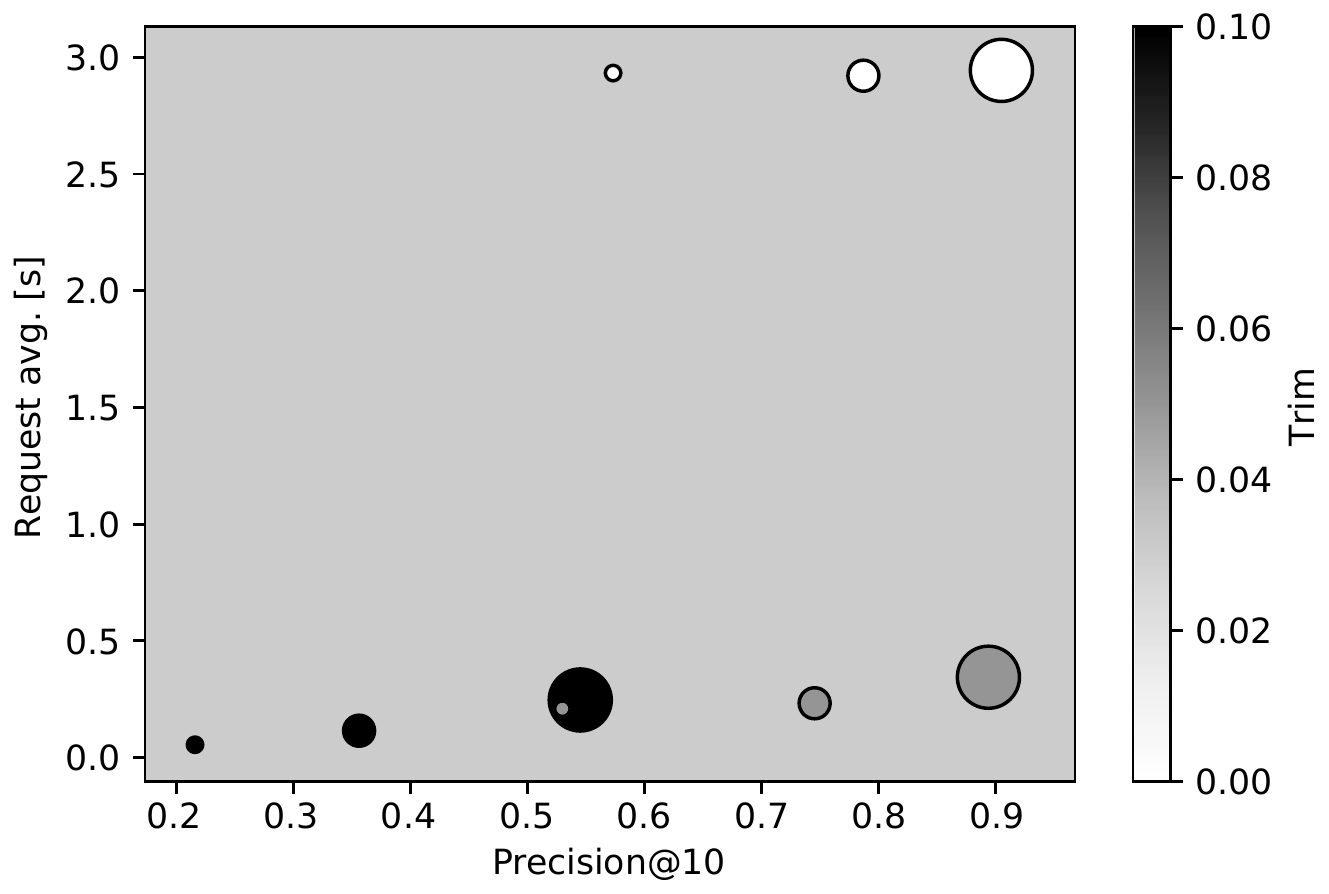}}
        \caption{Parallel querying in 4~parallel queues, 32~queries each (128~queries in total)}
        \label{fig:results:req-time-avg:paral4}
    \end{subfigure}
    \caption{The impact of value filtering and the number of retrieved search
             results on the average request time.
             The \textit{x-axis} shows the \textit{average precision},
             the \textit{y-axis} the \textit{average request time}.
             The \textit{size} of the points indicates the \textit{number of retrieved results}:
                large = \ES page size 320,
                medium = page size 80,
                small = 20.
             The \textit{color} of the points indicates the \textit{trim thresholds}:
                white = 0.00 (no filtering),
                gray  = 0.05,
                black = 0.10.}
    \label{fig:results:req-time-avg}
\end{figure*}

Figure~\vref{fig:results:quality} illustrates the impact of feature value filtering and the number of retrieved search candidates from \ES{} (page size) on its accuracy.
It can be seen that avg.\ diff.\ decreases logarithmically with the page size.
The results improve all the way up to 640 search results (the maximum value we have
tried), which is expected as this increases the size of the subset~$E$ that is 
consequently ordered (re-ranked) in phase~2 with the precise but more costly exact algorithm. 
Increasing the size of $E$ increases the chance of the inclusion of relevant 
results.

The shape of the curve suggests that there would only be a slight improvement in accuracy, and this would be at the cost of a substantial drop in performance.
The impact of including only a limited number of 
features with the highest absolute value (see Section~\vref{sec:hp-filtering}), is rather low. This is an excellent result with regard to performance as it means we may effectively sparsify the query vector with very little impact on search quality. 
We observe little difference between no filtering (searching by all 400 encoded 
features) and trimming to only the 90~best query vector values. However, 
trimming to as low as 6~values results in a significant increase in avg.\ diff.
See Figure~\vref{fig:results:quality:avg-diff}.

Our methods scores above the MLT baseline in all of the followed metrics, even with aggressive high-pass filtering and with a small page size.

We expect that similar setup parameters will work similarly, at least for general multi-topic text datasets. Its behaviour for a dramatically different dataset, such as images instead of texts, or without normalized feature ranges, cannot be directly inferred and remains to be investigated in our future work.
However, we expect our speed optimization methods to be applicable in some form, with the concrete parameters to be validated on the particular dataset and algorithmic setup.

\subsection{Speed Evaluation}
\label{sec:results:speed}

Selected results of our speed evaluation are summarized in Table~\vref{tab:results:speed}.
For clarity, we selected only the configurations using 400 LSA features and 48~\ES{} shards, as this 
setup turned out to provide the optimal performance on the \ES{} cluster and dataset we used according to the quality evaluation (see Section~\vref{sec:eval:quality}), where the same parameters were used.

\begin{table*}[tb]
    \renewcommand{\arraystretch}{1.15}
    \tabcolsep2.65dd
    \centering
\caption{Results of speed evaluation using 400 LSA features, batches of 128
         queries and 48 \ES{} shards.
         Column \textit{Parallel q.} shows the number of parallel queries used together,
         \textit{Trim} is the threshold for high-pass filtering of the features,
         \textit{Page} is the number of vectors retrieved from \ES for each query
           (see Section~\vref{sec:hp-filtering}),
         \textit{ES avg./std.} is the average and standard deviation of the
           number of seconds per request \ES took,
         \textit{Request avg./std.} is the average and deviation of the number
           of seconds per request including processing overheads,
         \textit{Total time} is the total number of seconds for processing all requests,
         \textit{Vec.\ size avg./std.} is the average/deviation of the number of
           values in query vectors that passed high-pass filtering.}
\vspace*{-.25\baselineskip}
\small\tabcolsep3.6dd
\bgroup\let~\cipherphantom
\begin{tabular}{c c c c c c c c c c c@{\,}}
    \toprule
\rotw{Parallel q.} &
\rotw{Trim} &
\rotw{Page} &
\rotw{ES avg.\ [s]} &
\rotw{ES std.\ [s]} &
\rotw{Request avg.\ [s]} &
\rotw{Request std.\ [s]} &
\rotw{Total time [s]} &
\rotw{Vec.\ size avg.} &
\rotw{Vec.\ size std.} \\
    \midrule
 ~1 & 0.00 & ~20 &  1.1949 & 0.0737 & ~1.2418 & 0.2864 & 160.760 & 400.0000 & ~0.0000 \\
 ~1 & 0.00 & ~80 &  1.1656 & 0.0710 & ~1.1829 & 0.0713 & 153.246 & 400.0000 & ~0.0000 \\
 ~1 & 0.00 & 320 &  1.2066 & 0.0893 & ~1.2494 & 0.0894 & 161.783 & 400.0000 & ~0.0000 \\
 ~1 & 0.05 & ~20 &  0.0935 & 0.0202 & ~0.1013 & 0.0204 & ~14.521 & ~93.5781 & 20.9268 \\
 ~1 & 0.05 & ~80 &  0.0935 & 0.0214 & ~0.1119 & 0.0228 & ~15.898 & ~91.5781 & 22.9822 \\
 ~1 & 0.05 & 320 &  0.1635 & 0.1014 & ~0.2085 & 0.1032 & ~28.360 & ~87.2500 & 20.6908 \\
 ~1 & 0.10 & ~20 &  0.0241 & 0.0057 & ~0.0320 & 0.0057 & ~~5.574 & ~19.3516 & ~4.4222 \\
 ~1 & 0.10 & ~80 &  0.0259 & 0.0059 & ~0.0435 & 0.0061 & ~~7.107 & ~19.0469 & ~4.3027 \\
 ~1 & 0.10 & 320 &  0.0940 & 0.0966 & ~0.1383 & 0.0971 & ~19.357 & ~18.0703 & ~4.7863 \\
 ~4 & 0.00 & ~20 &  2.9093 & 0.3943 & ~2.9325 & 0.3989 & ~94.750 & 400.0000 & ~0.0000 \\
 ~4 & 0.00 & ~80 &  2.8842 & 0.2968 & ~2.9211 & 0.3004 & ~94.548 & 400.0000 & ~0.0000 \\
 ~4 & 0.00 & 320 &  2.8621 & 0.2897 & ~2.9439 & 0.2938 & ~95.290 & 400.0000 & ~0.0000 \\
 ~4 & 0.05 & ~20 &  0.1919 & 0.0491 & ~0.2094 & 0.0531 & ~~7.212 & ~89.3516 & 21.9351 \\
 ~4 & 0.05 & ~80 &  0.2027 & 0.0525 & ~0.2327 & 0.0563 & ~~7.948 & ~93.2422 & 20.6166 \\
 ~4 & 0.05 & 320 &  0.2583 & 0.0983 & ~0.3442 & 0.1025 & ~11.538 & ~86.3203 & 24.1855 \\
 ~4 & 0.10 & ~20 &  0.0422 & 0.0080 & ~0.0550 & 0.0078 & ~~2.253 & ~18.8047 & ~4.4547 \\
 ~4 & 0.10 & ~80 &  0.0703 & 0.0708 & ~0.1151 & 0.0884 & ~~4.174 & ~19.5547 & ~4.2625 \\
 ~4 & 0.10 & 320 &  0.1547 & 0.1093 & ~0.2468 & 0.1161 & ~~8.411 & ~17.8750 & ~4.4459 \\
 16 & 0.00 & ~20 & 11.5664 & 3.2998 & 11.6019 & 3.3018 & ~94.480 & 400.0000 & ~0.0000 \\
 16 & 0.00 & ~80 & 11.5408 & 2.6209 & 11.6033 & 2.6262 & ~94.535 & 400.0000 & ~0.0000 \\
 16 & 0.00 & 320 & 11.5144 & 3.7843 & 11.7623 & 3.7645 & ~95.112 & 400.0000 & ~0.0000 \\
 16 & 0.05 & ~20 &  0.7870 & 0.1988 & ~0.8116 & 0.2020 & ~~6.896 & ~88.4141 & 21.6681 \\
 16 & 0.05 & ~80 &  0.7253 & 0.2309 & ~0.8802 & 0.3051 & ~~7.463 & ~87.9063 & 23.1528 \\
 16 & 0.05 & 320 &  0.8511 & 0.2350 & ~1.0453 & 0.2491 & ~~9.332 & ~89.9141 & 23.0604 \\
 16 & 0.10 & ~20 &  0.1354 & 0.0182 & ~0.1660 & 0.0169 & ~~1.625 & ~18.4375 & ~3.9484 \\
 16 & 0.10 & ~80 &  0.1845 & 0.0859 & ~0.2613 & 0.0891 & ~~2.400 & ~18.7656 & ~4.8404 \\
 16 & 0.10 & 320 &  0.4181 & 0.1442 & ~0.6213 & 0.2065 & ~~5.416 & ~18.0703 & ~4.3807 \\
    \bottomrule
\end{tabular}
\egroup
\label{tab:results:speed}
\end{table*}

The speed of the native \ES MLT search is summarized in 
Table~\vref{tab:results:speed:native-es}. The speed is comparable 
to our method when high-pass filtering is involved.

\begin{table}[tb]
  \caption{Results of speed evaluation using the native \ES More Like This (MLT) 
         search (no parallel queries) using 48~shards.
         \textit{\texttt{max\_query\_terms}} is the maximum number of query 
         terms per query that were selected by \ES.
         \textit{ES avg./std.} is the average and standard deviation of the
           number of seconds per request \ES took.}
  \vspace*{-.25\baselineskip}
  \renewcommand{\arraystretch}{1.15}
  \tabcolsep5dd\centering\small
  \begin{tabular}{c c c c}
    \toprule
    System & \texttt{max\_query\_terms} & ES avg. [s] & ES std. [s] \\
    \midrule
    MLT & \hphantom{0}17 & 0.0468 & 0.0233 \\
    MLT & \hphantom{0}25 & 0.0595 & 0.0270 \\
    MLT & \hphantom{0}40 & 0.0745 & 0.0322 \\
    MLT & \hphantom{0}90 & 0.1090 & 0.0490 \\
    MLT &            400 & 0.1458 & 0.0978 \\
    \bottomrule
\end{tabular}
\label{tab:results:speed:native-es}
\end{table}

Figure~\vref{fig:results:req-time-avg} displays the impact of value filtering 
and the number of retrieved search results on the average request time.
Comparing consecutive queries (Figure~\vref{fig:results:req-time-avg:single})
with four parallel queries on the same \ES cluster configuration
(Figure~\vref{fig:results:req-time-avg:paral4})
shows that at the expense of doubling the response time, we are able to answer four requests in parallel.

The best results in Figure~\vref{fig:results:req-time-avg} are located in the
bottom right corner where the precision is high and the response time is low.
For our dataset and algorithm (LSA with 400~features), the best overall
results are represented by the largest gray dots, i.e.\ retrieving
320 vectors from \ES while filtering the query vector to roughly 90~values via
trimming with a threshold of~0.05.

To achieve the optimal results, we suggest retrieving as large a set of candidates 
(\ES page size, $E$) as the response-time constraints allow, as the page size 
seems to have significantly lower influence on the response time compared to trimming, while having a significant positive effect on accuracy.

Our experiments were done on the Wikipedia dataset. Wikipedia is a multi-topic dataset~-- articles are on a wide variety of different topics using different keywords (names of people, places and things, etc) and notation (text only articles, articles on mathematics using formulae, articles on chemistry using different formulae, etc) are included. This provides enough room for the machine learning algorithms to build features that reflect these unique markers of particular topics and makes particular features significantly irrelevant for particular documents in the dataset.

For general multi-topic text datasets, we recommend trimming features values by their absolute value below 5\% of the maximum (i.e.\ between $-0.05$ and 0.05 in our experiments). Trimming more feature tokens decreases the precision with almost no influence on the response times, while keeping more feature tokens in the index has almost no positive effect on the precision but slows down the search significantly.
\vfill  

\section{Conclusions}
In this paper we have demonstrated a novel method for the conversion of semantic vectors into a set of string `feature tokens' that can be subsequently indexed in a standard inverted-index-based fulltext search engine, such as \ES.

Two techniques of feature tokens filtering were demonstrated to further significantly speed up the search process, with an acceptably low impact on the quality of the results.

Using \ES MLT on document texts as a baseline, our method
performs better than the baseline on all the followed metrics. With sufficient
query vector feature reduction, our method is faster than MLT.
With moderate query vector feature
reduction, we can achieve excellent approximation of the gold standard while
being only marginally slower than the MLT.

An important conclusion from our experiments is that the search speed can be improved even with filtering the query vectors alone and without the need to trim index vectors.
A pleasant practical consequence of this finding is that a vector search engine based on our proposed scheme could allow the users to define the filtering parameters dynamically, at search time rather than at indexing time. 
In this way, we let the users choose the approximation trade-off between the search speed and accuracy, i.e.\ use weaker filtering parameters for searches where accuracy is critical, and more aggressive filtering where speed is critical.

In our future work we will focus on the validation of our techniques on different types of data (such as images or audio data) and different text representations (such as doc2vec) in specific domains (such as question answering).

\subsubsection*{Acknowledgments}
Funding by TA ČR Omega grant TD03000295 is gratefully acknowledged.

\brokenpenalty10000


\begin{thebibliography}{}
\expandafter\ifx\csname natexlab\endcsname\relax\def\natexlab#1{#1}\fi

\bibitem[{Blei et~al.(2003)Blei, Ng, Jordan, and Lafferty}]{blei03lda}
David~M. Blei, Andrew~Y. Ng, Michael~I. Jordan, and John Lafferty. 2003.
\newblock \href{http://dl.acm.org/citation.cfm?id=944919.944937}{{Latent
  Dirichlet Allocation}}.
\newblock {\em Journal of Machine Learning Research\/} 3:993--1022.
\newblock
  \href{http://dl.acm.org/citation.cfm?id=944919.944937}{http://dl.acm.org/citation.cfm?id=944919.944937}.

\bibitem[{Boytsov(2017)}]{ir:boytsov2017}
Leonid Boytsov. 2017.
\newblock \href{http://boytsov.info/pubs/proposal\_boytsov.pdf}{{Efficient and
  Accurate Non-Metric $k$-NN Search with Applications to Text Matching}}.
\newblock Thesis proposal, School of Computer Science, Carnegie Mellon
  University.
\newblock
  \href{http://boytsov.info/pubs/proposal\_boytsov.pdf}{http://boytsov.info/pubs/proposal\_boytsov.pdf}.

\bibitem[{Deerwester et~al.(1990)Deerwester, Dumais, Furnas, Landauer, and
  Harshman}]{nlp:LSA1990}
Scott Deerwester, Susan~T. Dumais, George~W. Furnas, Thomas~K. Landauer, and
  Richard Harshman. 1990.
\newblock {Indexing by Latent Semantic Analysis}.
\newblock {\em Journal of the American Society for Information Science\/}
  41(6):391--407.

\bibitem[{Digout et~al.(2004)Digout, Nascimento, and Coman}]{ir:Digout2004}
Christian Digout, Mario~A. Nascimento, and Alexandru Coman. 2004.
\newblock \href{https://doi.org/10.1007/978-3-540-24571-1\_73}{Similarity
  search and dimensionality reduction: Not all dimensions are equally useful}.
\newblock In YoonJoon Lee, Jianzhong Li, Kyu-Young Whang, and Doheon Lee,
  editors, {\em Proc. of Database Systems for Advanced Applications: 9th Int.
  Conf., DASFAA 2004, Jeju Island, Korea, March 17--19, 2003\/}. Springer,
  pages 831--842.
\newblock
  \href{https://doi.org/10.1007/978-3-540-24571-1\_73}{https://doi.org/10.1007/978-3-540-24571-1\_73}.

\bibitem[{Gionis et~al.(1999)Gionis, Indyk, and
  Motwani}]{ir:conf/vldb/GionisIM99}
Aristides Gionis, Piotr Indyk, and Rajeev Motwani. 1999.
\newblock \href{http://www.vldb.org/conf/1999/P49.pdf}{Similarity search in
  high dimensions via hashing}.
\newblock In {\em VLDB\,'99, Proceedings of 25th International Conference on
  Very Large Data Bases, September 7--10, 1999, Edinburgh, Scotland, {UK}\/}.
  pages 518--529.
\newblock
  \href{http://www.vldb.org/conf/1999/P49.pdf}{http://www.vldb.org/conf/1999/P49.pdf}.

\bibitem[{Gormley and Tong(2015)}]{Gormley:2015:EDG:2904394}
Clinton Gormley and Zachary Tong. 2015.
\newblock {\em {Elasticsearch: The Definitive Guide}\/}.
\newblock O'Reilly Media, Inc., 1st edition.

\bibitem[{Le and Mikolov(2014)}]{ml:LeMikolov2014}
Quoc~V. Le and Tomas Mikolov. 2014.
\newblock \href{http://arxiv.org/abs/1405.4053}{Distributed representations of
  sentences and documents}.
\newblock {\em CoRR\/} abs/1405.4053.
\newblock
  \href{http://arxiv.org/abs/1405.4053}{http://arxiv.org/abs/1405.4053}.

\bibitem[{Manning et~al.(2008)Manning, Raghavan, and
  Sch\"{u}tze}]{ir:Manning:2008:IIR}
Christopher~D. Manning, Prabhakar Raghavan, and Hinrich Sch\"{u}tze. 2008.
\newblock {\em {Introduction to Information Retrieval}\/}.
\newblock Cambridge University Press, New York, NY, USA.

\bibitem[{Mikolov et~al.(2013)Mikolov, Sutskever, Chen, Corrado, and
  Dean}]{ml:mikolov2013distributed}
Tomas Mikolov, Ilya Sutskever, Kai Chen, Greg~S Corrado, and Jeff Dean. 2013.
\newblock Distributed representations of words and phrases and their
  compositionality.
\newblock In {\em Advances in Neural Information Processing Systems\/}. pages
  3111--3119.

\bibitem[{Rygl et~al.(2016)Rygl, Sojka, Růžička, and
  Řehůřek}]{ir:raslanscaletext2016}
Jan Rygl, Petr Sojka, Michal Růžička, and Radim Řehůřek. 2016.
\newblock
  \href{https://nlp.fi.muni.cz/raslan/2016/paper08-Rygl\_Sojka\_etal.pdf}{{ScaleText:
  The Design of a Scalable, Adaptable and User-Friendly Document System for
  Similarity Searches: Digging for Nuggets of Wisdom in Text}}.
\newblock In Aleš Horák, Pavel Rychlý, and Adam Rambousek, editors, {\em
  Proceedings of the Tenth Workshop on Recent Advances in Slavonic Natural
  Language Processing, RASLAN 2016\/}. Tribun EU, Brno, pages 79--87.
\newblock
  \href{https://nlp.fi.muni.cz/raslan/2016/paper08-Rygl\_Sojka\_etal.pdf}{https://nlp.fi.muni.cz/raslan/2016/paper08-Rygl\_Sojka\_etal.pdf}.

\bibitem[{Salton and Buckley(1988)}]{ml:SaltonBuckley1988}
Gerard Salton and Chris Buckley. 1988.
\newblock Term-weighting approaches in automatic text retrieval.
\newblock {\em Information Processing and Management\/} 24:513--523.

\bibitem[{Salton et~al.(1975)Salton, Wong, and Yang}]{ir:Salton1975}
Gerard Salton, Anita Wong, and Chung-Shu Yang. 1975.
\newblock \href{https://doi.org/10.1145/361219.361220}{A vector space model for
  automatic indexing}.
\newblock {\em Communications of the ACM\/} 18(11):613--620.
\newblock
  \href{https://doi.org/10.1145/361219.361220}{https://doi.org/10.1145/361219.361220}.

\bibitem[{Weber and B{\"o}hm(2000)}]{ir:Weber2000}
Roger Weber and Klemens B{\"o}hm. 2000.
\newblock \href{https://doi.org/10.1007/3-540-46439-5\_2}{Trading quality for
  time with nearest-neighbor search}.
\newblock In Carlo Zaniolo, Peter~C. Lockemann, Marc~H. Scholl, and Torsten
  Grust, editors, {\em Proc. of Advances in Database Technology --- EDBT 2000:
  7th Int. Conf. on Extending Database Technology Konstanz, Germany, March
  27--31, 2000\/}. Springer, pages 21--35.
\newblock
  \href{https://doi.org/10.1007/3-540-46439-5\_2}{https://doi.org/10.1007/3-540-46439-5\_2}.

\bibitem[{Weber et~al.(1998)Weber, Schek, and Blott}]{ir:Weber1998}
Roger Weber, Hans-J\"{o}rg Schek, and Stephen Blott. 1998.
\newblock \href{http://dl.acm.org/citation.cfm?id=645924.671192}{A quantitative
  analysis and performance study for similarity-search methods in
  high-dimensional spaces}.
\newblock In {\em Proc. of the 24rd International Conference on Very Large Data
  Bases\/}. Morgan Kaufmann Publishers Inc., San Francisco, CA, USA, VLDB '98,
  pages 194--205.
\newblock
  \href{http://dl.acm.org/citation.cfm?id=645924.671192}{http://dl.acm.org/citation.cfm?id=645924.671192}.

\bibitem[{Zezula et~al.(2006)Zezula, Amato, Dohnal, and Batko}]{ir:Zezula2006}
Pavel Zezula, Giuseppe Amato, Vlastislav Dohnal, and Michal Batko. 2006.
\newblock {\em {Similarity Search: The Metric Space Approach}\/}, volume~32 of
  {\em Advances in Database Systems\/}.
\newblock Springer.

\end{thebibliography}

\end{document}